\documentclass[lettersize,journal]{IEEEtran}
\usepackage{amsmath,amsfonts}
\usepackage{algorithm}
\usepackage{algpseudocode}
\usepackage{array}
\usepackage{textcomp}
\usepackage{stfloats}
\usepackage{url}
\usepackage{verbatim}
\usepackage{graphicx}
\usepackage{cite}
\usepackage{color}
\usepackage{bbm}  
\usepackage{subfigure}

\usepackage{booktabs} 
\usepackage{multirow} 
\usepackage{ulem}

\usepackage{hyperref}
\hypersetup{
	colorlinks=true,
	linkcolor=blue,
	citecolor=blue,
	urlcolor=blue
}

\begin{document}

\title{miMamba: EEG-based Emotion Recognition with Multi-scale Inverted Mamba Models}

\author{Xin Zhou, Dawei Huang, Xiaojiang Peng, \textit{Senior Member, IEEE}, Lijun Yin, \textit{Fellow, IEEE} 

\thanks{Xin Zhou and Lijun Yin are with the Department of Computer Science, State University of New York at Binghamton, Binghamton, NY 13902 USA. E-mail: (xzhou11, lyin)@binghamton.edu}

\thanks{Dawei Huang and Xiaojiang Peng (Corresponding author) are with the College of Big Data and Internet, Shenzhen Technology University, Shenzhen, 518118, China. E-mail: huangdawei2023@email.szu.edu.cn, pengxiaojiang@sztu.edu.cn}
}

\markboth{Journal of \LaTeX\ Class Files,~Vol.~14, No.~8, August~2021}%
{Shell \MakeLowercase{\textit{et al.}}: A Sample Article Using IEEEtran.cls for IEEE Journals}


\maketitle

\begin{abstract}
EEG-based emotion recognition holds significant potential in the field of brain-computer interfaces. A key challenge lies in extracting discriminative spatiotemporal features from electroencephalogram (EEG) signals. Existing studies often rely on domain-specific time-frequency features and analyze temporal dependencies and spatial characteristics separately, neglecting the interaction between local-global relationships and spatiotemporal dynamics. To address this, we propose a novel network called Multi-Scale Inverted Mamba (MS-iMamba), which consists of Multi-Scale Temporal Blocks (MSTB) and Temporal-Spatial Fusion Blocks (TSFB). Specifically, MSTBs are designed to capture both local details and global temporal dependencies across different scale subsequences. The TSFBs, implemented with an inverted Mamba structure, focus on the interaction between dynamic temporal dependencies and spatial characteristics. The primary advantage of MS-iMamba lies in its ability to leverage reconstructed multi-scale EEG sequences, exploiting the interaction between temporal and spatial features without the need for domain-specific time-frequency feature extraction. Experimental results on the DEAP, DREAMER, and SEED datasets demonstrate that MS-iMamba achieves classification accuracies of 94.86\%, 94.94\%, and 91.36\%, respectively, using only four-channel EEG signals, outperforming state-of-the-art methods.
\end{abstract}


\section{Introduction}
\IEEEPARstart{E}{motion} recognition is pivotal for enhancing human-computer interaction and intelligent systems. Accurately identifying emotional states enables systems to respond more appropriately to human needs, thereby improving interaction naturalness and efficiency. EEG, as a non-invasive method for physiological signal acquisition, offers superior temporal resolution and continuity compared to other signals like facial expressions or voice, allowing real-time capture of human's emotional dynamics. In practical applications, EEG-based emotion recognition can facilitate mental health monitoring and diagnostic support by detecting abnormal patterns related to emotional disorders, providing objective indicators for clinical diagnosis \cite{jafari2023emotion}.

Extracting and analyzing discriminative spatiotemporal features from EEG signals is a challenging task due to the brain's complex spatial topology. Traditional approaches often involve manual extraction of domain-specific time-frequency features such as differential entropy (DE) \cite{shi2013differential, song2018eeg}, power spectral density (PSD) \cite{lin2010eeg, zhang2021sparsedgcnn}, and functional connectivity \cite{li2019eeg}. While these methods have advanced EEG emotion recognition, they are time-consuming, require extensive domain knowledge, and often lose valuable temporal information by compressing long time series into single single eigenvalue.

To address these limitations, deep learning methods have gained prominence for their end-to-end capabilities. For example, Cui et al. \cite{cui2022eeg} utilized gated recurrent units combined with minimal class confusion for emotion recognition. Feng \cite{feng2022eeg} and Li \cite{li2020exploring} integrated attention mechanisms into bidirectional long short-term memory (LSTM) modules to extract key temporal features from EEG sequences. Similarly, Du et al. \cite{du2020efficient} applied attention mechanisms with LSTM-generated feature vectors to automatically select appropriate EEG channels for emotion recognition. Other studies have framed physiological signal emotion recognition tasks as sequence-to-sequence multivariate time series prediction problems, employing advanced self-attention mechanisms to decompose signals into independent frequency and time-domain representations \cite{yangdecompose}. These approaches effectively capture useful temporal dependencies. Given that individual EEG time steps lack semantic meaning \cite{nie2022time}, the appropriateness of iterating or calculating mutual correlations among them is questionable. Inspired by this, we segment EEG signals into patches of different scales, aggregating time steps into subsequence-level patches to enhance local details and capture global relationships that single time points cannot provide.

Other deep learning methods have been utilized to construct spatial features from EEG signals, significantly enhancing emotion recognition accuracy. Typical spatial feature extraction methods include convolutional neural networks (CNNs) \cite{ngai2022emotion, tao2020eeg}. For example, Li et al \cite{li2022eeg}. employed a novel efficient convolutional block to reduce computational burden. Liu et al. \cite{liu20213dcann} proposed a model named  3-D Convolutional Attention Neural Network (3DCANN), which consists of spatiotemporal feature extraction modules and EEG channel attention weight learning modules. This model effectively captures dynamic relationships between multi-channel EEG signals and the internal spatial relationships within these signals. Recent studies have shown that graph convolutions can effectively utilize brain topological structures for emotion recognition. Lin et al. \cite{lin2023eeg} developed an improved graph convolution model combined with dynamic channel selection to simulate information transmission in the brain. This model combines the advantages of one-dimensional convolution and graph convolution, capturing intra-channel and inter-channel EEG features and further modeling inter-regional brain connectivity by adding functional connectivity. Feng et al \cite{feng2022eeg}. introduced a spatial graph convolution module that adaptively learns intrinsic connections between EEG channels using an adjacency matrix to extract spatial domain features. Additionally, researchers like Cui \cite{cui2020eeg} and Deng et al. \cite{deng2021sfe} explored the spatial information of adjacent and symmetrical channels from the perspective of whether EEG signals exhibit symmetrical emotional responses. These studies underscore the importance of understanding brain topological structures in EEG-based emotion recognition tasks.

The brain's complex structure results in EEG signals with time-varying spatial topology and temporal dependencies recorded through multiple electrodes. It is intuitive to use both temporal dependencies and spatial features as auxiliary information. For instance, a novel Attention-based Spatiotemporal Dual-Stream Fusion Network (ASTDF-Net) \cite{gong2023astdf} has been employed to learn a joint latent subspace to capture the coupled spatiotemporal information in EEG signals. Cheng and Feng et al. proposed a hybrid model combining a Spatial-Graph Convolutional Network (SGCN) module and an attention-enhanced bidirectional Long Short-Term Memory (LSTM) module \cite{feng2022eeg}, and later designed a hybrid network comprising a Dynamic Graph Convolution (DGC) module and a Temporal Self-Attention Representation (TSAR) module, integrating spatial topology and temporal information \cite{cheng2023hybrid}. Gong et al. \cite{gong2023eeg} used a novel Attention-based Convolutional Transformer Neural Network (ACTNN), effectively integrating key spatial, spectral, and temporal information of EEG signals and cascading CNNs with transformers for emotion recognition tasks. Shen et al. \cite{shen2022multiscale} utilized multi-scale temporal self-attention modules to learn temporal continuity information while employing dynamic graph convolution modules to capture spatial functional relationships between different EEG electrodes. Although these integrated models consider both temporal and spatial features, they typically use two separate branches to extract spatiotemporal information, lacking adequate interaction between them.

Given these challenges, this article proposes a spatiotemporal fusion network called Multi-Scale Inverted Mamba (MS-iMamba), combining Multi-Scale Temporal Blocks (MSTB) and Temporal-Spatial Fusion Blocks (TSFB). The primary advantage of the proposed MS-iMamba is its ability to simultaneously leverage local details and global relationships in EEG signals without the need for complex statistical feature extraction, enhancing emotion recognition performance through adequate spatiotemporal dependency interactions. Specifically, MSTB divides EEG signals into multiple scale patches, using small-scale patches to capture fine local details and coarse global relationships, thereby utilizing complementary predictive capabilities in multi-scale observations. TSFB embeds the temporal dimension rather than the spatial dimension of reconstructed multi-scale EEG signals into a token and uses a Selective State-Space Model (SSM) to model spatiotemporal information. This mechanism fully integrates the spatiotemporal dependencies of both modules to enhance EEG emotion recognition.

The main contributions of this study are as follows:

\begin{itemize}
	
	\item A plug-and-play MSTB is designed, which considers local details and global relationships without requiring traditional domain-specific time-frequency statistical feature extraction, providing a promising perspective for time dependency modeling in EEG emotion recognition.
	
	\item The proposed TSFB reflects on the modeling approach of the spatiotemporal characteristics of EEG signals, adequately considering the interaction between temporal dependencies and spatial features, offering a method that simultaneously integrates temporal and spatial features for EEG emotion recognition.
	
	\item Intra-subject and inter-subject experiments were conducted on three public datasets: DEAP, DREAMER, and SEED. The experimental results demonstrate that the proposed MS-iMamba outperforms various state-of-the-art methods using only four-channel EEG data.
\end{itemize}

The remaining sections of this article are organized as follows. Section \ref{sec2} reviews the related work on multi-scale and spatiotemporal representation learning. Section \ref{sec3} presents the pipeline of MS-iMmaba. Section \ref{sec4} details the procedure of the conducted experiments and experimental results. A more in-depth discussion is provided in Section \ref{sec5}. Finally, the study is concluded in Section \ref{sec6}.

\section{Related Work}\label{sec2}
\subsection{Multi-Scale Representation Learning}
Representing data in fine granularity has been widely adopted in time series prediction \cite{nie2022time} and computer vision fields \cite{dosovitskiy2020image, zhu2024vision}. EEG emotion recognition is essentially a time series prediction task, and considering an effective sequence representation approach is necessary. Nie et al.\cite{nie2022time} argued that for time series data, single-point data lacks clear semantic information unlike words, making the computation of single-time-step correlations debatable. In natural language processing, it is also more effective to symbolize words in a sentence rather than each letter \cite{devlin2018bert, schuster2012japanese}. This approach of aggregating single-point data into patches has been validated effectively in time series prediction tasks \cite{wang2024timemixer}. For instance, Wu et al. \cite{wu2022timesnet} addressed the limitations of 1-D time series by segmenting the sequence into short and long periods representations. These representations were embedded into the columns and rows of a 2-D tensor to capture inter- and intra-periodic variations, respectively, allowing easy modeling of 2-D variations using 2-D convolutional kernels. Chen et al. \cite{chen2024pathformer} highlighted the difficulty of capturing features across multiple scales when modeling time series with limited or fixed scales. Their proposed Pathformer model achieved multi-scale modeling by integrating time resolution and time distance, dividing the time series into different time resolutions and performing dual attention mechanisms at each scale to capture global correlations and local details as temporal dependencies.

Multi-scale representation learning has also been applied to EEG signal processing. Wang et al. \cite{wang2020linking} proposed the Multi-Scale Convolutional Neural Network-Dynamic Graph Convolutional Network (AMCNN-DGCN) model to avoid the cumbersome manual feature extraction process. Jiang et al.\cite{jiang2023emotion} designed a novel Attention Mechanism-Based Multi-Scale Feature Fusion Network (AM-MSFFN) that considers high-level features at different scales to enhance the model's generalization capability across different subjects. To extract a comprehensive range of multi-class features from multi-channel EEG time series for accurate understanding of brain activity, Li et al. \cite{li2020multi} introduced a Multi-Scale Attention Mechanism Fusion Convolutional Neural Network (MS-AMF), which extracts spatiotemporal multi-scale features from signals representing multiple brain regions and employs a dense fusion strategy to retain maximum information flow. These studies underscore the importance of multi-scale representation learning in EEG temporal modeling and demonstrate its potential in the field of emotion recognition.

\subsection{Spatiotemporal Representation Learning}
In the field of multivariate time series prediction, the fusion of spatiotemporal features has become a popular strategy for improving prediction accuracy. Numerous scholars have focused on effectively integrating temporal continuity with spatial correlations \cite{bai2020adaptive, sesti2021integrating}. Grigsby et al. proposed Spacetimeformer \cite{grigsby2021long}, which transforms multivariate time series problems into a spatiotemporal sequence format. In this approach, each input token represents the value of a single variable at a given time step, allowing simultaneous learning of temporal and spatial relationships. Other works have modeled spatiotemporal relationships by transforming one-dimensional or multidimensional sequence data into tensors \cite{chen2020low, sharma2022classification}. Jin et al. \cite{jin2023graph} demonstrated that traditional methods, which process multichannel EEG signals into one-dimensional graphical features, limit the expressive capability of emotion recognition models. To address this issue, they introduced the G2G module, which transforms one-dimensional graphical data into two-dimensional grid data through numerical relationship encoding, using deep models like ResNet for subsequent tasks. Li et al. \cite{li2024multi} employed dilated causal convolutional neural networks to extract nonlinear relationships between different time frames and used feature-level fusion to merge features from multiple channels, exploring potential complementary information between different views to enhance feature representation.

Recently, the integration of attention mechanisms and graph neural networks for EEG spatiotemporal modeling has gained increasing attention. Cheng and Feng have conducted extensive research in this direction. Initially, they proposed a model combining a Spatial Graph Convolution Network (SGCN) module with an attention-enhanced bidirectional Long Short-Term Memory (LSTM) module. This model's main advantage is its consideration of each brain region's biological topology, extracting representative spatiotemporal features from multiple EEG channels \cite{feng2022eeg}. They subsequently designed a hybrid network comprising a Dynamic Graph Convolution (DGC) module and a Temporal Self-Attention Representation (TSAR) module, incorporating representative knowledge of spatial topology and temporal context into EEG emotion recognition tasks\cite{cheng2023hybrid}. Recently, they equipped the Dense Graph Convolutional Network (DGC) with Joint Cross-Attention (JCA) for multimodal emotion recognition tasks, termed DG-JCA \cite{cheng2024dense}. However, Zeng et al.\cite{zeng2023transformers} demonstrated that single-layer linear models unexpectedly outperformed complex Transformer-based models in time series prediction tasks. Liu et al. \cite{liu2023itransformer} reflected on this result, suggesting that for multivariate sequence data, points on different channels at the same time step record entirely different physical meanings or events, making embedding them into tokens inappropriate. They proposed the Inverted Transformer (iTransformer), which treats independent time series as tokens and captures multivariate correlations through self-attention to leverage spatiotemporal mutual information.

These methods share a common goal of revealing and utilizing the spatiotemporal features of time series data from different perspectives to achieve higher prediction accuracy. Each method has its specific application scenarios and advantages, but they all highlight the importance of spatiotemporal feature fusion in current research, providing a wealth of technical options and research directions for the field of EEG emotion recognition.

\begin{figure}[t]
	\centering
	\includegraphics[width=\linewidth]{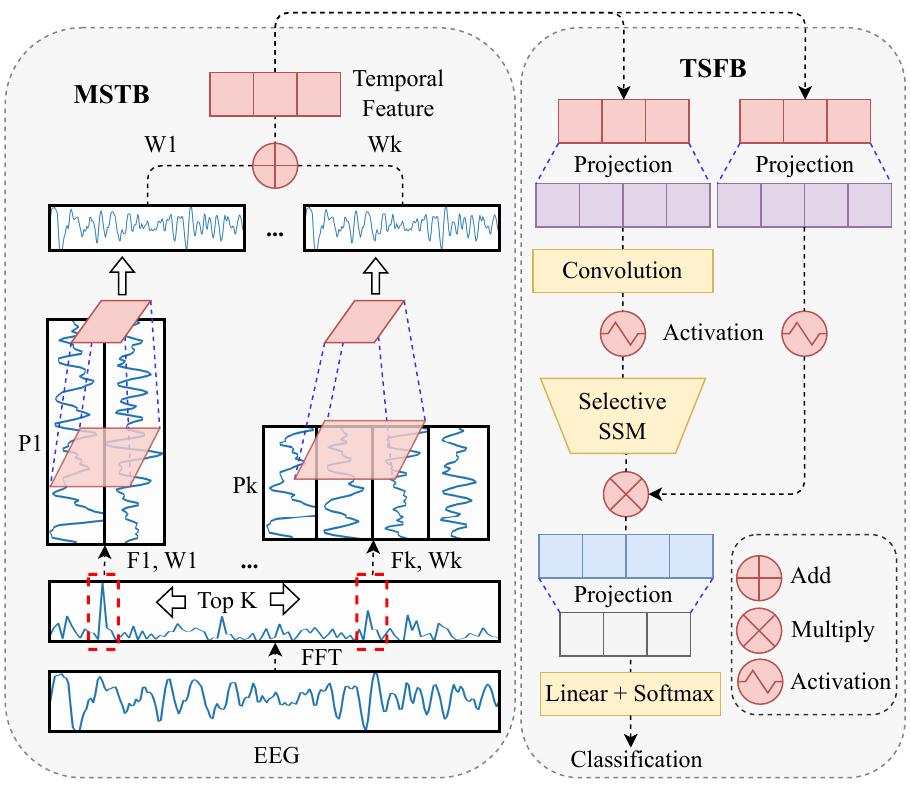}
	\caption{Architecture of the MS-iMamba network for EEG emotion recognition. The network comprises two main modules: the Multi-Scale Temporal Block (MSTB) and the Temporal-Spatial Fusion Block (TSFB). The MSTB extracts multi-scale representations by converting the EEG signal into different frequency domain components and reshaping them into 2-D patches. These patches capture both local and global dependencies through convolution operations. The TSFB then integrates temporal and spatial information by embedding multiple time steps of the same channel into tokens, enabling effective feature extraction through the iMamba module, which combines a reversed embedding mechanism with a selective spatial state model (SSM).}
	\label{fig.ms_imamba}
\end{figure}

\section{Method}\label{sec3}
In this section, we formalize the MS-iMamba network for EEG emotion recognition. As illustrated in Figure \ref{fig.ms_imamba}, the network consists of two primary modules: the Multi-Scale Temporal Block (MSTB) and the Temporal-Spatial Fusion Block (TSFB). Each module will be discussed in detail below.

\subsection{Notations and Definitions}

Let the EEG signals of each subject be represented by a 3-D matrix \( S \in \mathbb{R}^{M \times T \times C} \), where \( M \), \( T \), and \( C \) denote the number of trials, sampling points, and channels, respectively. The matrix \( S \) is segmented into \( N \) samples of length \( L \) using a non-overlapping sliding window (thus, \( T = N \times L \)). The segmented EEG samples are denoted as \( \mathcal{I} = \{ (X_{ij}, Y_{ij}) \mid i = 1, 2, \ldots, M; j = 1, 2, \ldots, N \} \), where \( X_{ij} \in \mathbb{R}^{L \times C} \) and \( Y_{ij} \in \mathbb{R} \) represent the ground-truth label corresponding to \( X_{ij} \). For the same trial, all \( N \) segments share the same label. Each segmented sample is denoted as \( X_{\text{1D}} := X_{ij} \). The goal of EEG emotion recognition is to predict \( Y_{ij} \) given \( X_{\text{1D}} \).

\subsection{Multi-Scale Temporal Block (MSTB)}
\subsubsection{Multi-Scale Representation}

Let \( X_{\text{1D}} \) denote an EEG signal of length \( L \) with \( C \) channels. Before representing this signal in a multi-scale format, we need to determine the patch sizes. To achieve this, we transform the original EEG signal into the frequency domain for analysis. Specifically, as shown in Equation \ref{eq.1}:

\begin{equation}
	\label{eq.1}
	\begin{gathered}
		A=\mathcal{A}\left(F F T\left(X_{1 \mathrm{D}}\right)\right), \\
		\left\{f_1, f_2, \ldots, f_k\right\}=\operatorname{argTop}_k(A), \\
		p_i=\left\lceil L / f_i\right\rceil, \quad i \in\{1, \ldots, k\},
	\end{gathered}
\end{equation}

\noindent where \( FFT \) denotes the Fast Fourier Transform (FFT), and $\mathcal{A}$ represents the amplitude calculation for each frequency. Since high-frequency regions often contain significant noise, we select only the top \( k \) frequencies with the highest amplitudes to avoid interference . The selected frequencies \(\{f_1, f_2, \ldots, f_k\}\) correspond to periods and amplitudes \(\{p_1, p_2, \ldots, p_k\}\) and \(\{A_{f_1}, A_{f_2}, \ldots, A_{f_k}\}\), respectively. The periods \(\{p_1, p_2, \ldots, p_k\}\) are used as the patch sizes for segmenting the EEG signal.

As illustrated in the left part of Figure \ref{fig.ms_imamba}, the original EEG signal is transformed into the frequency domain using FFT, with the red dashed boxes indicating the \( k \) frequencies with the highest amplitudes. We then calculate the weights for each frequency using Equation \ref{eq.2}:

\begin{equation}
	\label{eq.2}
	\begin{gathered}
		W_{f_i} = \{W_{f_1}, \ldots, W_{f_k}\} = \text{Softmax}(A_{f_1}, \ldots, A_{f_k}).
	\end{gathered}
\end{equation}

\noindent Next, the signal \( X_{\text{1D}} \) is segmented into patches of varying sizes and reshaped into a 2-D format, as shown in Equation \ref{eq.3}:

\begin{equation}
	\label{eq.3}
	X^i_{\text{2D}} = \text{Reshape}_{p_i, f_i}(\text{Padding}(X_{\text{1D}})), \quad i \in \{1, \ldots, k\},
\end{equation}

\noindent where the padding operation ensures the original sequence can be divided into integer patches. The reshaped EEG signal is represented in a multi-scale format, \( X^i_{\text{2D}} \in \mathbb{R}^{p_i \times f_i \times C} \), which denotes the \( i \)-th reshaped time series based on period \( p_i \). The vertical and horizontal directions represent intra-patch and inter-patch variations, respectively. These variations capture local details and global relationships. Consequently, we obtain a set of 2-D tensors derived from different patches \(\{X^1_{\text{2D}}, X^2_{\text{2D}}, \ldots, X^k_{\text{2D}}\}\). This transformation facilitates capturing information at various distances, with larger \( p_i \) capturing longer temporal dependencies. Additionally, the reshaped tensors allow for efficient feature extraction using convolutional operations.

\subsubsection{Multi-Scale Perception}

The reshaped tensors are processed by the Multi-Scale Perception $(\mathcal{MSP})$ module, as shown in Equation \ref{eq.4}:

\begin{equation}
	\label{eq.4}
	{X}^i_{\text{1D}} = \text{Reshape}_{1, p_i \times f_i}({\mathcal{MSP}}(X^i_{\text{2D}})), \quad i \in \{1, \ldots, k\}.
\end{equation}

\noindent In this module, convolutional kernels of different sizes are employed. This mechanism allows the module to simultaneously perceive variations within the same patch and across patches with the same phase. After the convolution operations, we reshape the 2-D tensors back to the 1-D form \({X}^i_{\text{1D}}\). To assign different levels of attention to features extracted from patches corresponding to different frequencies, we perform a weighted sum of these multi-scale signals to obtain the final reconstructed multi-scale representation, as shown in Equation \ref{eq.5}:

\begin{equation}
	\label{eq.5}
	X_{\text{1D}} = \sum_{i=1}^{k} W_{f_i} \times {X}{^i}_{\text{1D}}.
\end{equation}

\noindent This approach ensures that the features from various scales are effectively combined, enhancing the overall representation of the EEG signal for emotion recognition.

\begin{figure}[t]
	\centering
	\includegraphics[width=\linewidth]{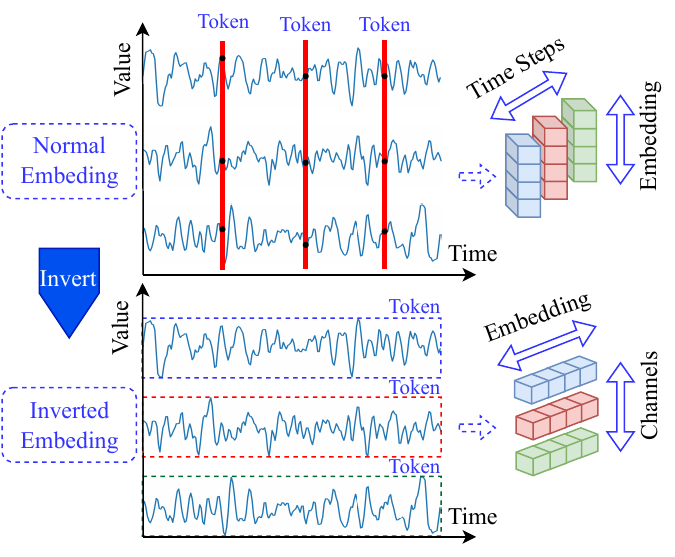}
	\caption{Comparison between normal and inverted embedding mechanism. The top part illustrates the conventional embedding approach, where data from different channels at the same time step are mapped into a single token. The bottom part depicts the reversed embedding method, where multiple time steps of the same channel are mapped into a single token.
	}
	\label{fig.embedding}
\end{figure}

\subsection{Temporal-Spatial Fusion Block (TSFB)}
\subsubsection{Inverted Embedding Representation}

After obtaining the multi-scale representation of the EEG signal, we consider the interaction of temporal and spatial information. Generally, conventional methods embed data from different channels at the same time step into a single token. As illustrated in the upper part of Figure \ref{fig.embedding}, the conventional embedding method places points from different electrodes, each representing completely different events and physical meanings, into the same token. Specifically, at a certain time point, some channel data might be at a peak while others are at a trough. Embedding them into the same token not only fails to reveal valuable information due to the narrow focus of a single time point but also represents misaligned events as a single token.

Therefore, we adopt an inverted embedding method, as shown in the lower part of Figure \ref{fig.embedding}. The inverted embedding method maps multiple time steps of the same channel into a single token. This event-driven representation not only considers information over longer time steps but also distinguishes data from different channels through separate tokens. The inverted embedding representation approach enhances the capacity to capture temporal dependencies and spatial relationships, ensuring a more comprehensive and meaningful interpretation of the EEG signals for emotion recognition.

This method is demonstrated through the following equations. Given a multi-scale EEG representation \(X_{\text{1D}}\), we reshape it to consider the temporal and spatial interactions:

\begin{equation}
	\label{eq.6}
	\hat{X}_{\text{1D}} = \text{Reshape}_{C,L}(X_{\text{1D}}).
\end{equation}

\noindent Here, \(\hat{X}_{\text{1D}}\) represents that temporal steps and channels are reorganized to reflect the inverted embedding structure. To capture the dynamic interactions between temporal and spatial features, we apply a SSM to \(\hat{X}_{\text{1D}}\). This inversion of the embedding representation and the fusion of temporal-spatial information using SSM enhance the ability to model the complex dependencies in EEG signals, leading to improved performance in emotion recognition tasks.

%
%

\subsubsection{iMamba}
After the inverted embedding operation, \(X_{1D} \in \mathbb{R}^{L \times C}\) is transformed into \(\hat{X}_{1D} \in \mathbb{R}^{C \times L}\). Next, we introduce the iMamba model, which consists of the inverted embedding mechanism and the SSM, specifically Mamba, to capture spatiotemporal correlations.

Mamba is inspired by continuous systems, mapping a 1-D sequence through a hidden state \(h(t) \in \mathbb{R}^N\) to \(x(t) \in \mathbb{R} \rightarrow y(t) \in \mathbb{R}\). As shown in Equation \ref{eq.8}, Mamba uses three parameter matrices \(\mathbf{A} \in \mathbb{R}^{d \times d}\), \(\mathbf{B} \in \mathbb{R}^{d \times 1}\), and \(\mathbf{C} \in \mathbb{R}^{1 \times d}\) (where \(d\) is the hidden dimension) to control this process. These parameters are analogous to the forget gate, input gate, and output gate mechanisms in LSTM. The parameter \(\mathbf{A}\) controls how much information is ignored, \(\mathbf{B}\) controls how the current input affects the hidden state, and \(\mathbf{C}\) controls the output flow of information:

\begin{equation}
	\label{eq.8}
	\begin{gathered}
		H'(t) = \mathbf{A}h(t) + \mathbf{B}x(t), \\
		\quad y(t) = \mathbf{C}h(t).
	\end{gathered}
\end{equation}

To adapt to discrete sequences, Mamba uses zero-order hold techniques, transforming \(\mathbf{A}\) and \(\mathbf{B}\) into their discrete versions via the time scale parameter \(\Delta\), as defined below:

\begin{equation}\label{eq.9}
	\begin{aligned}
		& \overline{\mathbf{A}}=\exp (\Delta \mathbf{A}), \\
		& \overline{\mathbf{B}}=(\Delta \mathbf{A})^{-1}(\exp (\Delta \mathbf{A})-\mathbf{I}) \cdot \Delta \mathbf{B}.
	\end{aligned}
\end{equation}

\noindent The discrete version is redefined as follows:

\begin{equation}
	\begin{aligned}
		h_t & =\overline{\mathbf{A}} h_{t-1}+\overline{\mathbf{B}} x_t, \\
		y_t & =\mathbf{C} h_t.
	\end{aligned}
\end{equation}

\noindent For parallel processing, Mamba computes the output using the following convolution form:

\begin{equation}
	\begin{aligned}
		\overline{\mathbf{K}} & =\left(\mathbf{C} \overline{\mathbf{B}}, \mathbf{C} \overline{\mathbf{A B}}, \ldots, \mathbf{C} \overline{\mathbf{A}}^{L-1} \overline{\mathbf{B}}\right), \\
		\mathbf{y} & =\mathbf{x} * \overline{\mathbf{K}},
	\end{aligned}
\end{equation}

\noindent where \(L\) is the length of input sequence \(\mathbf{x}\) and \(\hat{K}\) is the structured convolution kernel.

Due to the inverted embedding operation, iMamba can extract both temporal and spatial features from the input data, fully considering spatiotemporal interactions. iMamba receives the input \(\hat{X}_{1D} \in \mathbb{R}^{C \times L}\) and produces the output prediction \(\hat{Y_i}\) through the following calculation:

\begin{equation}
	\hat{Y_i} = f(\text{iMamba}(\hat{X}_{1D})),
\end{equation}

\noindent where \(f(\cdot)\) is a linear classifier consisting of a Linear layer and a softmax operation. The cross-entropy loss is computed as follows:

\begin{equation}
	\mathcal{L}_{c l s}=-\sum_{i=1}^{n} \mathbbm{1}_{\left[i = \hat{Y_i}\right]} \log \left(\hat{Y_i}\right),
\end{equation}

\noindent where \(n\) denotes the number of categories and $\mathbbm{1}_{[i = \hat{Y_i}]}$ equals 1 if the predicted class matches the true label, and 0 otherwise. Finally, the backpropagation algorithm is used to update the network parameters. The pseudocode of MS-iMamba is summarized in Algorithm \ref{Algorithm}.

\begin{algorithm}[th]
	\caption{MS-iMamba for EEG Emotion Recognition} 
	\label{Algorithm}
	\begin{algorithmic}[1]
		\Require EEG signal $S \in \mathbb{R}^{M \times T \times C}$
		\Ensure Predicted label $\hat{Y}$
		
		\State \textbf{Preprocessing:}
		\State Slice $S$ into non-overlapping windows to get samples $I = \{(X_{ij}, Y_{ij})\}$
		
		\State \textbf{Multi-Scale Temporal Block (MSTB):}
		\State Transform $X_{ij}$ to frequency domain using FFT
		\State $A = \mathcal{A}(\text{FFT}(X_{\text{1D}}))$
		\State Select top $k$ frequencies and their periods:
		\State $\{f_1, f_2, \ldots, f_k\} = \text{argTop}_k(A), \quad p_i = \lceil L / f_i \rceil$
		\State Calculate weights:
		\State $W_{f_i} = \text{Softmax}(A_{f_1}, A_{f_2}, \ldots, A_{f_k})$
		\State Reshape $X_{\text{1D}}$ into 2-D patches:
		\State $X^{i}_{2D} = \text{Reshape}_{p_i, f_i} (\text{Padding}(X_{\text{1D}}))$
		\State Apply multi-scale inception:
		\State ${X}^{i}_{1D} = \text{Reshape}_{1, p_i \times f_i} (\mathcal{MSP}(X^{i}_{2D}))$
		\State Combine multi-scale features:
		\State $X_{\text{1D}} = \sum_{i=1}^{k} W_{f_i} \times {X}^{i}_{1D}$
		
		\State \textbf{Temporal-Spatial Fusion Block (TSFB):}
		\State Reverse embedding to reshape $\hat{X}_{\text{1D}} \in \mathbb{R}^{C \times L}$
		
		\State \textbf{iMamba:}
		\State Apply iMamba to capture spatio-temporal correlation:
		\State $\hat{Y}_{i} = f(\text{iMamba}(\hat{X}_{\text{1D}}))$
		\State Calculate cross-entropy loss:
		\State $\mathcal{L}_{cls} = \sum_{i=1}^{n} \log(\hat{Y}_{i}) \cdot \mathbbm{1}[i = \hat{Y}_i]$
		\State Update network parameters using backpropagation.
		
	\end{algorithmic}
\end{algorithm}

\section{Experiment and Results Analysis}\label{sec4}

\subsection{Datasets}
DEAP \cite{koelstra2011deap}: The DEAP dataset comprises multimodal data collected from 32 participants. Each participant watched 40 music videos while 32-channel EEG data and 8-channel peripheral physiological signals were recorded. Participants rated the videos on a scale from 1 to 9 for valence, arousal, dominance, and liking. Each video contains 60 seconds of data (excluding a 3-second baseline signal), which was downsampled to 128 Hz and filtered using a 4-45 Hz band-pass filter. We classified each metric into high and low categories using a threshold of 5. To augment the dataset, we segmented each signal into 1-second non-overlapping segments. In our experiments, we used only the frontal polar region channels FP1, FP2, AF3, and AF4.

DREAMER \cite{katsigiannis2017dreamer}: The DREAMER dataset also contains multimodal data from 23 participants. Each participant watched 18 video clips (ranging from 65s to 393s, with an average duration of 199s), while 14-channel EEG and 2-channel Electrocardiograph (ECG) signals were recorded. Participants rated valence, arousal, and dominance on a scale from 1 to 5. The signals were sampled at 128 Hz and filtered to 4-45 Hz using a band-pass filter. The EEG signals were then segmented into 1-second non-overlapping segments to expand the dataset. For DREAMER, we used four channels from the frontal polar and frontal regions: AF3, AF4, F7, and F8. Each metric was classified into high and low categories using a threshold of 3.

SEED \cite{zheng2015investigating}: The SEED dataset includes data from 15 participants, with 62-channel EEG data collected according to the international 10-20 system. Each participant conducted three sessions approximately one week apart, during which they watched 15 different film clips (each lasting about 4 minutes). These films elicited positive, neutral, and negative emotions as experimental stimuli. The data were downsampled to 200 Hz and filtered to 0-75 Hz, then segmented into 1-second non-overlapping segments. Only the frontal polar region channels FP1, FP2, AF3, and AF4 were selected for our experiments. Finally, to mitigate data drift across different channels, Z-score normalization was applied to all three datasets.

\subsection{Training Protocol}
In our experiments, we employed two different paradigms: intra-subject and inter-subject paradigm. For the intra-subject paradigm, we evaluated each participant's data individually, using 80\% for training and 20\% for testing. For the inter-subject paradigm, we combined and shuffled the data from all participants, splitting it into training and testing sets in a 4:1 ratio. Due to the SEED dataset comprising data from three different sessions, which significantly impacts experimental results, we also used intra-session and inter-session evaluation methods. Our training configuration included a batch size of 32, the Adam optimizer with an initial learning rate of \(1 \times 10^{-3}\), and 10 epochs of training. An adaptive learning rate strategy was employed to reduce the learning rate as the training loss decreased. Other hyperparameters, such as the number of network layers and Top-k, were set to 1 and 2, respectively. All experiments were conducted on an Intel Xeon Silver 4210R CPU @ 2.40GHz (×2) and an NVIDIA RTX A6000 GPU.

\subsection{Baseline Model Selection}
For our benchmark model selection, we chose several representative methods to compare against our model under the same experimental settings. These models are sourced from the Time Series Library (TSlib\footnote{TSLib: \url{https://github.com/thuml/Time-Series-Library}}) and include the top three ranked models for classification tasks: TimesNet, Non-stationary Transformer (NTransformer), and Informer. Additionally, we included models characterized by linear structures and causal convolution structures, such as DLinear and TCN. Below is a brief introduction to these benchmark models:

\begin{itemize}
	
	\item iTransformer \cite{liu2023itransformer}: iTransformer addresses the shortcomings of traditional Transformers in modeling spatiotemporal information by proposing an inverted Transformer structure that better considers spatiotemporal relationships.
	
	\item DLinear \cite{zeng2023transformers}: DLinear decomposes sequences into periodic and trend components, achieving impressive results in various time series tasks using a straightforward linear structure, outperforming many complex Transformer models and their variants.
	
	\item TimesNet \cite{wu2022timesnet}: TimesNet employs a multi-scale strategy to transform time series from 1-D to 2-D format, capturing both intra-period and inter-period variations.
	
	\item NTransformer \cite{liu2022non}: This model designs non-stationary attention mechanisms to recover inherent non-stationary information in time dependencies through distinguishable attention learned from the raw sequences.
	
	\item Informer \cite{zhou2021informer}: Informer utilizes sparse attention and a self-distillation mechanism to reduce the computational complexity of attention maps to logarithmic levels.
	
	\item TCN \cite{bai2018empirical}: TCN introduces the concept of dilated causal convolutions, which are favored for expanding the receptive field without increasing computational burden.
	
\end{itemize}

\subsection{Intra-subject Experiment Results}

\begin{table}[ht]
	\centering
	\caption{Performance Comparison of Models on DEAP and DREAMER Datasets (Intra-subject)}
	\begin{tabular}{p{1.5cm}p{1.1cm}p{1.1cm}p{1.2cm}p{1.2cm}}
		\toprule
		Model Name & DEAP (valence)  & DEAP (arousal) & DREAMER (valence)  & DREAMER (arousal)  \\
		\hline
		iTransformer & 79.10\% & 81.35\% & 77.60\% & 79.78\% \\
		Dlinear & \uline{90.77}\% & \uline{91.89\%} & \uline{93.73}\% & \textbf{95.47\%} \\
		TimesNet & 87.32\% & 88.05\% & 84.69\% & 88.47\% \\
		NTransformer & 85.01\% & 87.01\% & 84.51\% & 86.25\% \\
		Informer & 87.27\% & 88.39\% & 86.48\% & 89.47\% \\
		TCN & 88.07\% & 89.24\% & 84.13\% & 88.63\% \\
		MS-iMamba & \textbf{94.69\%} & \textbf{95.03\%} & \textbf{94.54\%} & \uline{95.34\%} \\
		\bottomrule  
			
	\end{tabular}
	\label{tab1}
\end{table}

As shown in Table \ref{tab1}, MS-iMamba demonstrates outstanding performance on both the DEAP and DREAMER datasets, significantly outperforming other models in most metrics. Specifically, it achieves the highest accuracy in DEAP (valence) at 94.69\%, DEAP (arousal) at 95.03\%, and DREAMER (valence) at 94.54\%. It also achieves the second-highest accuracy in DREAMER (arousal) at 95.34\%, underscoring its robustness and effectiveness in emotion recognition tasks. This makes MS-iMamba an excellent choice for applications requiring high-precision valence and arousal detection from the DEAP and DREAMER datasets. Notably, the linear model DLinear also performs well in this context, second only to MS-iMamba, and even achieving the highest accuracy in DREAMER (arousal). Surprisingly, TCN surpasses several Transformer-based models, while TimesNet performs comparably to them.

\begin{table}[ht]
	\centering
	\caption{Performance Comparison of Models on SEED Dataset (Intra-subject, Inter-session and Intra-session)}
	\begin{tabular}{p{1.5cm}cccc}
		\toprule
		Model Name& Inter session & Session 1 & Session 2 & Session 3 \\
		\midrule
		iTransformer & 56.55\% & 60.18\% & 62.17\% & 54.28\% \\
		Dlinear & 64.51\% & \uline{85.80}\% & \uline{88.71\%} & \uline{81.07\%} \\
		TimesNet & \uline{70.21}\% & 79.79\% & 78.58\% & 68.93\% \\
		NTransformer & 65.10\% & 75.20\% & 76.84\% & 63.15\% \\
		Informer & 66.01\% & 79.15\% & 81.12\% & 68.84\% \\
		TCN & 67.47\% & 75.36\% & 73.31\% & 60.33\% \\
		MS-iMamba & \textbf{92.60\%} & \textbf{93.24\%} & \textbf{93.19\%} & \textbf{87.67\%} \\
		\bottomrule
	\end{tabular}
	\label{tab2}
\end{table}

Table \ref{tab2} presents the accuracy of different models across four sessions: inter-session, session 1, session 2, and session 3. MS-iMamba consistently maintains the highest accuracy in all sessions, demonstrating its strong performance and adaptability to various session conditions. The most significant improvement is observed in the inter-session scenario, where MS-iMamba outperforms the second-best model by approximately 22.39\%. This substantial advantage highlights MS-iMamba's exceptional ability to generalize across different session data. In specific session scenarios, MS-iMamba surpasses the second-best model by 7.44\%, 4.48\%, and 6.60\%, respectively. DLinear consistently ranks second in sessions 1, 2, and 3, indicating its reliability and effectiveness, although it lags noticeably in the inter-session scenario. TimesNet shows relatively high accuracy in the inter-session scenario (70.21\%), but its performance declines in subsequent sessions, indicating potential limitations in session-specific contexts. Other models, such as iTransformer, NTransformer, and TCN, exhibit lower and more variable performance across sessions, indicating less consistency compared to MS-iMamba and DLinear. Overall, MS-iMamba demonstrates superior performance across all scenarios, significantly outperforming other models, particularly under inter-session conditions. This consistent and robust performance makes MS-iMamba an exceptional model for tasks requiring high accuracy under different session conditions. DLinear emerges as a strong contender, particularly effective in single-session scenarios, but falls short in terms of generalization compared to MS-iMamba.

\subsection{Inter-subject Experiment Results}

\begin{table}[ht]
	\centering
	\caption{Performance Comparison of Models on DEAP and DREAMER Datasets (Inter-subject)}
	\begin{tabular}{p{1.5cm}p{1.1cm}p{1.1cm}p{1.2cm}p{1.2cm}}
		\toprule
		Model Name & DEAP (valence)  & DEAP (arousal) & DREAMER (valence)  & DREAMER (arousal)  \\
		\hline
		iTransformer & 63.82\% & 65.65\% & 64.19\% & 74.29\% \\
		Dlinear & 65.81\% & 66.28\% & 71.47\% & 78.21\% \\
		TimesNet & 67.58\% & 69.17\% & 66.39\% & 75.65\% \\
		NTransformer & 66.24\% & 68.76\% & 67.60\% & 75.75\% \\
		Informer & 65.31\% & 65.98\% & 67.84\% & 74.82\% \\
		TCN & \uline{75.80\%} & \uline{78.87\%} & \uline{72.24\%} & \uline{79.01\%} \\
		MS-iMamba & \textbf{86.04\%} & \textbf{85.94\%} & \textbf{81.90\%} & \textbf{87.04\%} \\
		\bottomrule
	\end{tabular}
	\label{tab3}
\end{table}

Table \ref{tab3} presents the performance of various models under inter-subject conditions on the DEAP and DREAMER datasets. Compared to the intra-subject paradigm, the inter-subject setting poses a greater challenge for model generalization. MS-iMamba consistently outperforms other models in both intra-subject and inter-subject conditions, demonstrating significant robustness and generalization capability. However, due to increased data variability, all models exhibit a performance drop when transitioning from intra-subject to inter-subject conditions. Despite its excellent performance in intra-subject scenarios, DLinear shows a notable decline in inter-subject settings, highlighting potential limitations in handling data from different subjects. TCN maintains relatively stable performance across both conditions, making it a reliable choice, albeit not the top-performing one.

\begin{table}[ht]
	\centering
	\caption{Performance Comparison of Models on SEED Dataset (Inter-subject, Inter-session, and Intra-session)}
	\begin{tabular}{p{1.5cm}cccc}
		\toprule
		Model Name& Inter session & Session 1 & Session 2 & Session 3 \\
		\midrule
		iTransformer & 38.76\% & 48.11\% & 47.80\% & 43.75\% \\
		Dlinear & 39.82\% & 44.88\% & 44.50\% & 42.66\% \\
		TimesNet & 47.39\% & 57.88\% & \uline{55.11\%} & \uline{51.59\%} \\
		NTransformer & 43.68\% & 53.04\% & 48.79\% & 45.95\% \\
		Informer & 43.46\% & 50.60\% & 47.61\% & 46.01\% \\
		TCN & \uline{68.25\%} & \uline{71.23\%} & 50.35\% & 46.57\% \\
		MS-iMamba & \textbf{86.10\%} & \textbf{93.71\%} & \textbf{94.54\%} & \textbf{89.70\%} \\
		\bottomrule
	\end{tabular}
	\label{tab4}
\end{table}

We also conducted inter-subject experiments on the SEED dataset, along with inter-session and intra-session experiments, with results presented in Table \ref{tab4}. MS-iMamba outperforms the second-best model by 17.85\%, 22.48\%, 40.43\%, and 38.11\% in inter-session, session 1, session 2, and session 3, respectively. Although the accuracy of MS-iMamba in the inter-session experiment decreases by 6.5\% compared to intra-subject conditions, its performance in intra-session experiments increases, while other models experience significant drops. Interestingly, while the DLinear model performs impressively in intra-subject experiments, it disappoints in inter-subject experiments, displaying the opposite pattern to TCN.

These results indicate that linear models are only suitable for scenarios with simple data structure distribution. Additionally, despite the multi-scale and inverted spatiotemporal structures used by TimesNet and iTransformer, their performance remains unsatisfactory. Overall, MS-iMamba consistently outperforms other models across nearly all datasets in both intra-subject and inter-subject experiments, showcasing superior robustness and generalization capabilities.

\begin{table}[ht]
	\centering
	\caption{Configurations of MS-iMamba variants used in the ablation experiments}
	\begin{tabular}{cccc}
		\toprule
		{Variants} & MSTB & Mamba & Inverted Embedding \\
		\midrule
		Multi-Scale & \checkmark &  &  \\
		Mamba &  & \checkmark &  \\
		Multi-Scale+Mamba & \checkmark & \checkmark &  \\
		iMamba &  & \checkmark & \checkmark \\
		Multi-Scale+iMamba & \checkmark & \checkmark & \checkmark \\

		\bottomrule
	\end{tabular}
	\label{tab5}
\end{table}

\begin{figure*}[htbp]
	\centering{
		\subfigure[{DEAP (valence)}]{
			\includegraphics[width=0.21\linewidth]{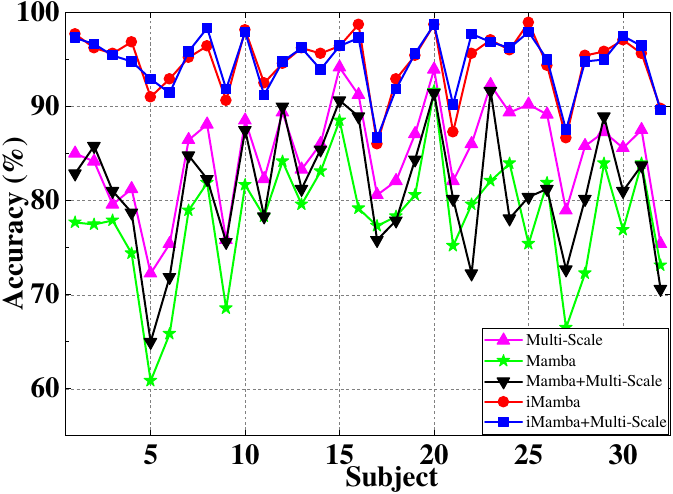}
			\label{fig.deap_v}
		}
		\quad
		\subfigure[{DEAP (arousal)}]{
			\includegraphics[width=0.21\linewidth]{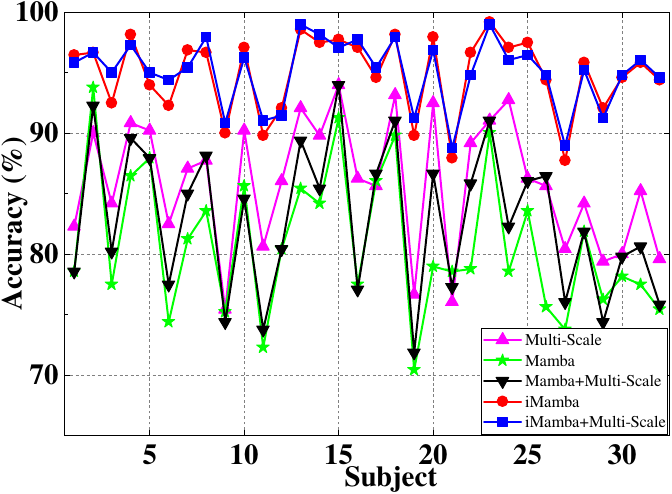}
			\label{fig.deap_a}
		}
		\quad
		\subfigure[{DREAMER (valence)}]{
			\includegraphics[width=0.21\linewidth]{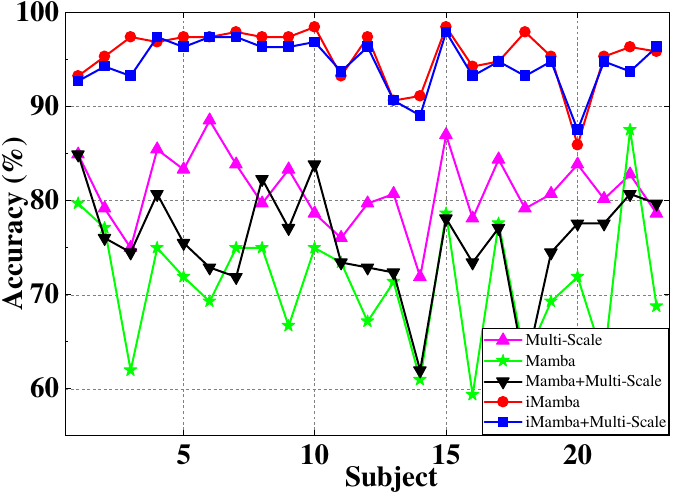}
			\label{fig.dreamer_v}
		}
		\quad
		\subfigure[{DREAMER (arousal)}]{
			\includegraphics[width=0.21\linewidth]{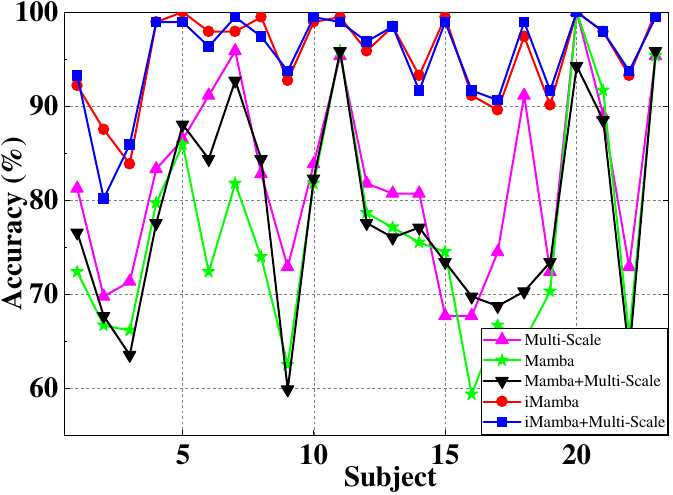}
			\label{fig.dreamer_a}
		}
	}

	\caption{
		Performance of different MS-iMamba variants on the DEAP and DREAMER datasets under intra-subject conditions.}
	\label{fig.deap_dreamer}
\end{figure*}

\begin{figure*}[htbp]
	\centering{
		\subfigure[{SEED (inter-session)}]{
			\includegraphics[width=0.21\linewidth]{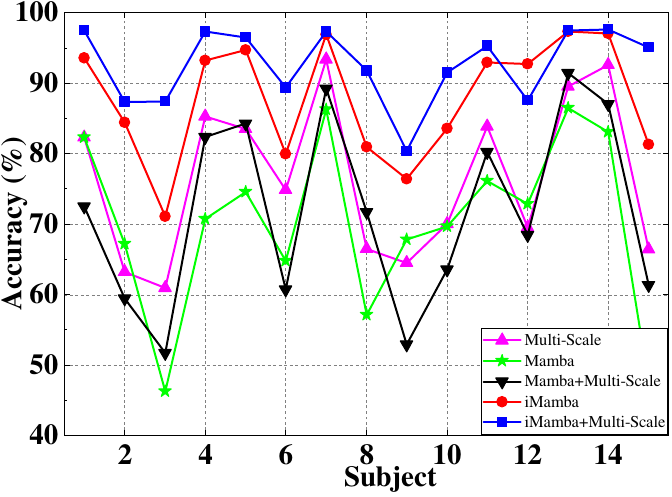}
			\label{fig.seed_inter}
		}
		\quad
		\subfigure[{SEED (session 1)}]{
			\includegraphics[width=0.21\linewidth]{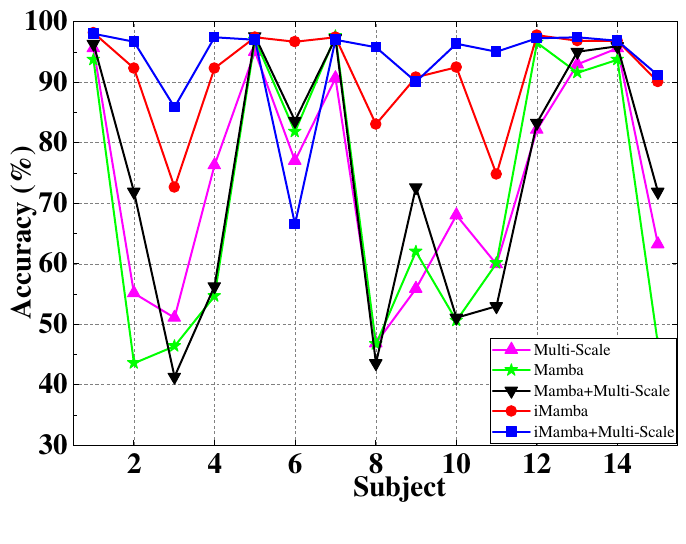}
			\label{fig.seed_s1}
		}
		\quad
		\subfigure[{SEED (session 2)}]{
			\includegraphics[width=0.21\linewidth]{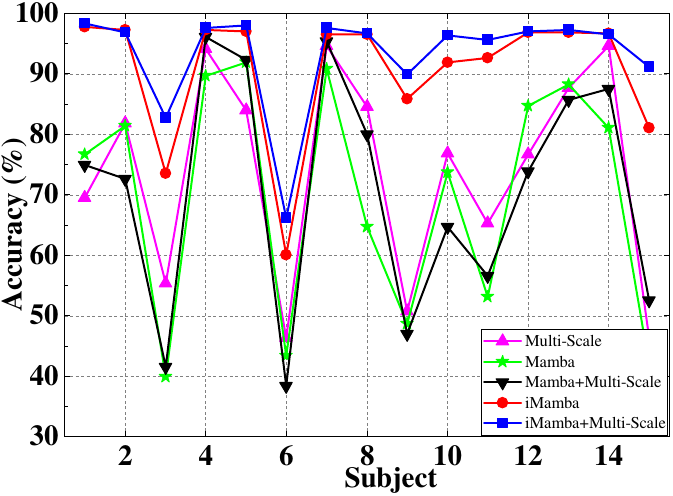}
			\label{fig.seed_s2}
		}
		\quad
		\subfigure[{SEED (session 3)}]{
			\includegraphics[width=0.21\linewidth]{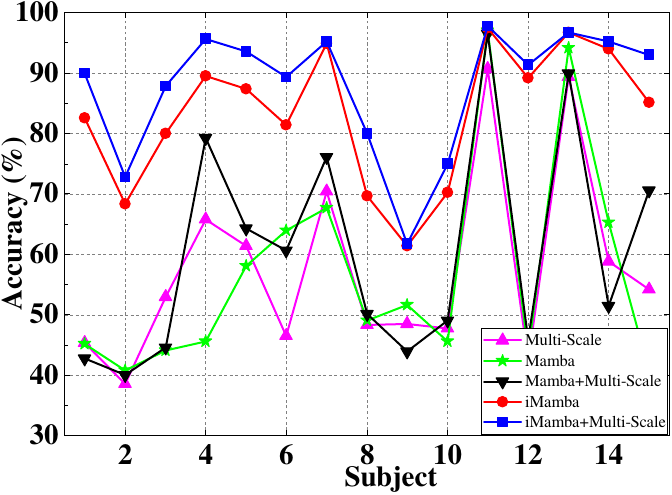}
			\label{fig.seed_s3}
		}
	}

	\caption{
		Performance of different MS-iMamba variants on the SEED dataset across four session modes under intra-subject conditions.}
	\label{fig.seed}
\end{figure*}

\subsection{Ablation Study}
To validate the effectiveness of each component in MS-iMamba, we conducted ablation experiments using five different configurations, as shown in Table \ref{tab5}. We compared the performance of these five variants under both intra-subject and inter-subject conditions.

\subsubsection{Intra-subject Results}
We visualized the performance of the five variants across three datasets, with the results illustrated in Figures \ref{fig.deap_dreamer} and \ref{fig.seed}. Figures \ref{fig.deap_v} and \ref{fig.deap_a} show the accuracy of each individual’s data on valence and arousal in the DEAP and DREAMER datasets, respectively. From the figures, we observe that the Mamba (green) performs the worst, while the Mamba with MSTB (black) shows slight improvement. However, both are outperformed by the variant using only MSTB (pink). This indicates that MSTB can effectively extract temporal features for emotion classification but does not integrate well with Mamba. The iMamba variant with inverted embedding (red) exhibits significant improvement, closely approaching the performance of MS-iMamba. These results suggest that in intra-subject scenarios on the DEAP and DREAMER datasets, using inverted embedding to consider spatiotemporal interactions is more beneficial than using multi-scale features. Figure \ref{fig.seed} displays the performance of these variants in four different session modes on the SEED dataset. The results indicate that Mamba shows significant improvement with the addition of MSTB and inverted embedding, with the latter providing a more substantial effect. These findings validate the effectiveness of MSTB and inverted embedding across all three datasets.

\begin{table*}[ht]
	\centering
	\caption{Performance of different MS-iMamba variants on the DEAP, DREAMER, and SEED datasets under inter-subject conditions.}
	\begin{tabular}{cccccccccc}
		\toprule

		\multirow{2}{*}{Datasets}  & DEAP   & DEAP  & DREAMER   & DREAMER  &SEED   & SEED  & SEED   & SEED &\multirow{2}{*}{Mean}   \\
		
		  & (valence)  &  (arousal) &  (valence)  &  (arousal) & (Inter session)  &  (Session 1) &  (Session 2)  &  (Session 3)  &\\
		
		\hline
		Multi-Scale & 71.20\% & 72.29\% & 70.81\% & 77.84\% & 57.90\% & 75.87\% & 73.89\% & 63.92\% &70.47\%
		\\
		Mamba & 67.70\% & 71.28\% & 65.91\% & 74.76\% & 53.36\% & 67.25\% & 67.42\% & 53.10\% &65.10\%
		 \\
		Mamba+Multi-Scale & 69.51\% & 71.71\% & 66.21\% & 74.62\% & 56.49\% & 66.54\% & 70.25\% & 59.33\% &66.83\%
		 \\
		iMamba & \uline{84.48\%} & \uline{85.51\%} & \uline{77.10\%} & \uline{81.21\%} & \uline{82.07\%} & \uline{91.05\%} & \uline{89.99\%} & \uline{77.98\%} &\uline{83.67\%}
		\\
		iMamba+Multi-Scale & \textbf{86.04\%} & \textbf{85.94\%} & \textbf{81.90\%} & \textbf{87.04\%} & \textbf{86.10\%} & \textbf{93.71\%} & \textbf{94.54\%} & \textbf{89.70\%} &\textbf{88.12\%}
		\\
		\bottomrule

	\end{tabular}
	\label{tab6}
\end{table*}

\subsubsection{Inter-subject Results}
We evaluated the performance of different MS-iMamba variants in inter-subject scenarios on the three datasets. Table \ref{tab6} shows that the combinations Mamba+Multi-Scale and iMamba+Multi-Scale, equipped with MSTB, achieve average accuracy improvements of 1.73\% and 4.45\%, respectively, compared to their counterparts without MSTB (Mamba and iMamba). The variants with inverted embedding (iMamba and iMamba+Multi-Scale) show average accuracy increases of 18.57\% and 21.29\%, respectively, compared to the Mamba and Mamba+Multi-Scale variants. The combined use of both mechanisms in iMamba+Multi-Scale (i.e., MS-iMamba) results in average accuracy improvements of 17.65\% and 23.02\% over the single-use Multi-Scale and Mamba variants. Overall, in inter-subject conditions, using MSTB, inverted embedding, or their combination leads to improved recognition performance on the DEAP, DREAMER, and SEED datasets.

\begin{table}[ht]
	\centering
	\caption{Average classification accuracy compared with the state-of-the-art methods on three datasets}
	\begin{tabular}{p{2.2cm}p{0.4cm}cp{0.8cm}p{1.2cm}p{0.7cm}}
		\toprule
		Models& Input & Channels & DEAP& DREAMER & SEED \\
		\midrule
		CRAM\cite{li2022eeg} & Raw & All & 85.78\% & 92.65\% & - \\
		JO-CapsNet\cite{li2022multi} & Raw  & All & 94.36\% &  & -\\
		DGCNN\cite{song2018eeg} & PSD  & All & 91.07\% & 85.39\% & 90.40\%\\
		IAG\cite{song2020instance} & PSD  & All & - & 90.89\% & -\\
		V-IAG\cite{song2021variational} & PSD  & All & - & \uline{92.96\%} & -\\
		EESCN\cite{xu2024eescn} & DE  & All & \uline{94.81\%} & - & -\\
		TAE\cite{cheng2024novel} & DE  & 30\% & 66.29\% & - & -\\
		ATDD-LSTM\cite{du2020efficient} & DE  & All & - & - & \uline{91.08\%}\\
		CSGNN\cite{lin2023eeg} & DE  & 20\% & 83.39\% & - & 83.93\%\\
		Ours & Raw  & 4 & \textbf{94.86\%} & \textbf{94.94\%} & \textbf{91.36\%}\\
		
		\bottomrule
	\end{tabular}
	\label{tab7}
\end{table}

\subsection{Comparison with State-of-the-Art Methods}

We compared MS-iMamba against state-of-the-art methods, and Table \ref{tab7} presents the average classification accuracies of various models on the DEAP, DREAMER, and SEED datasets. The input feature types include raw data (Raw), power spectral density (PSD), and differential entropy (DE). Among models utilizing feature extraction and all-channel EEG data, EESCN \cite{xu2024eescn}, V-IAG \cite{song2021variational}, and ATDD-LSTM \cite{du2020efficient} achieved the highest accuracies on the three datasets, with 94.81\%, 92.96\%, and 91.08\%, respectively. TAE \cite{cheng2024novel}, masking 70\% of the data and using the remaining 30\%, reached an accuracy of 66.29\% on DEAP, while CSGNN \cite{lin2023eeg}, retaining only 20\% of the channels, achieved accuracies of 83.39\% on DEAP and 83.93\% on SEED. These models performed poorly under conditions of incomplete data. In contrast, MS-iMamba, without using handcrafted features and relying on just four channels, achieved or exceeded the performance of these models. This demonstrates that our model effectively utilizes limited channel information to achieve high-precision classification. MS-iMamba consistently outperformed the state-of-the-art models across all datasets, highlighting the advanced nature and efficacy of our feature extraction and classification algorithms.

\section{Discussion}\label{sec5}
In this study, we designed MS-iMamba for EEG-based emotion recognition, incorporating two main components: Multi-Scale Temporal Blocks (MSTB) and Temporal-Spatial Feature Blocks (TSFB). MSTB and TSFB are utilized to capture multi-scale temporal features and spatiotemporal interactions, respectively. We replaced traditional manual time-frequency feature extraction with MSTB and introduced a novel approach to handle spatiotemporal information. The proposed model was compared with numerous advanced models, demonstrating its effectiveness. This section delves deeper into the discussion.

We employed three popular public datasets, DEAP, DREAMER and SEED, and used only four-channel EEG signals from the frontal polar region as inputs. This choice was based on two considerations. First, previous research indicates that emotion-related EEG signals are predominantly found in the prefrontal lobe and lateral temporal lobe of the brain \cite{li2024multi, gong2023eeg, lin2023eeg}. Second, the frontal polar region has less hair, reducing the likelihood of EEG signal interference from hair. Achieving high recognition accuracy with fewer EEG channels is a valuable exploration. Additionally, manual feature extraction requires specific domain knowledge and can disrupt the temporal characteristics of the original EEG signals, adding to the workload and potentially diminishing the dataset's usability.

With the rise of deep learning, self-attention mechanisms have garnered attention across various fields. However, our experiments revealed that Transformer-based models did not perform as expected with limited channels. Properly considering spatiotemporal characteristics can not only enhance recognition performance but also improve the interpretability of the EEG's temporal dependencies and spatial topology. Our two plug-and-play modules, MSTB and TSFB, are suited for different scenarios. From the experimental results, TSFB offered more significant benefits than MSTB. In simple data distribution scenarios, MSTB's improvement was minimal, whereas in complex environments, MSTB proved to be a valuable addition. Combining both modules enhances the model's generalization and robustness.

While MS-iMamba achieved impressive results using fewer channels, there are still several limitations. For instance, under the same experimental configuration, MS-iMamba's performance in cross-subject and cross-session scenarios was suboptimal. Given the challenges in acquiring large-scale EEG data, predicting unknown subjects' emotional categories using data from a few subjects remains challenging. However, our work suggests a potential method to preserve the data scale in EEG emotion recognition. In the future, we will continue exploring effective use of limited or incomplete data to improve MS-iMamba's performance in complex scenarios.

\section{Conclusion}\label{sec6}
This study introduces MS-iMamba, a novel model designed for EEG-based emotion recognition, integrating Multi-Scale Temporal Blocks (MSTB) and Temporal-Spatial Feature Blocks (TSFB). Our approach effectively captures multi-scale temporal features and spatiotemporal interactions, offering a robust alternative to traditional manual feature extraction methods. Comprehensive experiments conducted on three widely-used public datasets, DEAP, DREAMER, and SEED, demonstrate that MS-iMamba outperforms state-of-the-art models and achieves higher classification accuracy with fewer EEG channels.

Our results highlight the model's robustness and generalization capabilities. Notably, the combination of MSTB and TSFB enhances the model's performance, providing significant improvements over individual components. Despite the challenges in cross-subject and cross-session contexts, MS-iMamba's ability to achieve high accuracy with limited data channels underscores its potential for practical applications in real-world settings.

While MS-iMamba shows promise, it also faces limitations, particularly in handling the variability inherent in cross-subject and cross-session data. Future research will focus on further optimizing the model to handle these complexities and exploring the use of limited or incomplete data to enhance performance in more challenging scenarios.

In conclusion, MS-iMamba represents a significant advancement in EEG-based emotion recognition, offering a scalable, high-accuracy solution that balances the need for fewer data channels with robust performance. This work lays a foundation for future exploration in efficient and effective emotion recognition using EEG, with potential applications across various domains requiring precise emotional state detection.

\section*{Acknowledgments}
This work is partially supported by the National Natural Science Foundation of China (62176165), the Stable Support Projects for Shenzhen Higher Education Institutions (20220718110918001), the Natural Science Foundation of Top Talent of SZTU (GDRC202131), the Basic and Applied Basic Research Project of Guangdong Province (2022B1515130009), and the Special subject on Agriculture and Social Development, Key Research and Development Plan in Guangzhou (2023B03J0172).

\bibliographystyle{IEEEtran}
\normalem
\bibliography{root}{}

\begin{thebibliography}{10}
\providecommand{\url}[1]{#1}
\csname url@samestyle\endcsname
\providecommand{\newblock}{\relax}
\providecommand{\bibinfo}[2]{#2}
\providecommand{\BIBentrySTDinterwordspacing}{\spaceskip=0pt\relax}
\providecommand{\BIBentryALTinterwordstretchfactor}{4}
\providecommand{\BIBentryALTinterwordspacing}{\spaceskip=\fontdimen2\font plus
\BIBentryALTinterwordstretchfactor\fontdimen3\font minus \fontdimen4\font\relax}
\providecommand{\BIBforeignlanguage}[2]{{%
\expandafter\ifx\csname l@#1\endcsname\relax
\typeout{** WARNING: IEEEtran.bst: No hyphenation pattern has been}%
\typeout{** loaded for the language `#1'. Using the pattern for}%
\typeout{** the default language instead.}%
\else
\language=\csname l@#1\endcsname
\fi
#2}}
\providecommand{\BIBdecl}{\relax}
\BIBdecl

\bibitem{jafari2023emotion}
M.~Jafari, A.~Shoeibi, M.~Khodatars, S.~Bagherzadeh, A.~Shalbaf, D.~L. Garc{\'\i}a, J.~M. Gorriz, and U.~R. Acharya, ``Emotion recognition in eeg signals using deep learning methods: A review,'' \emph{Computers in Biology and Medicine}, p. 107450, 2023.

\bibitem{shi2013differential}
L.-C. Shi, Y.-Y. Jiao, and B.-L. Lu, ``Differential entropy feature for eeg-based vigilance estimation,'' in \emph{2013 35th Annual International Conference of the IEEE Engineering in Medicine and Biology Society (EMBC)}.\hskip 1em plus 0.5em minus 0.4em\relax IEEE, 2013, pp. 6627--6630.

\bibitem{song2018eeg}
T.~Song, W.~Zheng, P.~Song, and Z.~Cui, ``Eeg emotion recognition using dynamical graph convolutional neural networks,'' \emph{IEEE Transactions on Affective Computing}, vol.~11, no.~3, pp. 532--541, 2018.

\bibitem{lin2010eeg}
Y.-P. Lin, C.-H. Wang, T.-P. Jung, T.-L. Wu, S.-K. Jeng, J.-R. Duann, and J.-H. Chen, ``Eeg-based emotion recognition in music listening,'' \emph{IEEE Transactions on Biomedical Engineering}, vol.~57, no.~7, pp. 1798--1806, 2010.

\bibitem{zhang2021sparsedgcnn}
G.~Zhang, M.~Yu, Y.-J. Liu, G.~Zhao, D.~Zhang, and W.~Zheng, ``Sparsedgcnn: Recognizing emotion from multichannel eeg signals,'' \emph{IEEE Transactions on Affective Computing}, vol.~14, no.~1, pp. 537--548, 2021.

\bibitem{li2019eeg}
P.~Li, H.~Liu, Y.~Si, C.~Li, F.~Li, X.~Zhu, X.~Huang, Y.~Zeng, D.~Yao, Y.~Zhang \emph{et~al.}, ``Eeg based emotion recognition by combining functional connectivity network and local activations,'' \emph{IEEE Transactions on Biomedical Engineering}, vol.~66, no.~10, pp. 2869--2881, 2019.

\bibitem{cui2022eeg}
H.~Cui, A.~Liu, X.~Zhang, X.~Chen, J.~Liu, and X.~Chen, ``Eeg-based subject-independent emotion recognition using gated recurrent unit and minimum class confusion,'' \emph{IEEE Transactions on Affective Computing}, vol.~14, no.~4, pp. 2740--2750, 2022.

\bibitem{feng2022eeg}
L.~Feng, C.~Cheng, M.~Zhao, H.~Deng, and Y.~Zhang, ``Eeg-based emotion recognition using spatial-temporal graph convolutional lstm with attention mechanism,'' \emph{IEEE Journal of Biomedical and Health Informatics}, vol.~26, no.~11, pp. 5406--5417, 2022.

\bibitem{li2020exploring}
C.~Li, Z.~Bao, L.~Li, and Z.~Zhao, ``Exploring temporal representations by leveraging attention-based bidirectional lstm-rnns for multi-modal emotion recognition,'' \emph{Information Processing \& Management}, vol.~57, no.~3, p. 102185, 2020.

\bibitem{du2020efficient}
X.~Du, C.~Ma, G.~Zhang, J.~Li, Y.-K. Lai, G.~Zhao, X.~Deng, Y.-J. Liu, and H.~Wang, ``An efficient lstm network for emotion recognition from multichannel eeg signals,'' \emph{IEEE Transactions on Affective Computing}, vol.~13, no.~3, pp. 1528--1540, 2020.

\bibitem{yangdecompose}
Z.~Yang and H.~Cao, ``Decompose time and frequency dependencies: Multivariate time series physiological signal emotion recognition.''

\bibitem{nie2022time}
Y.~Nie, N.~H. Nguyen, P.~Sinthong, and J.~Kalagnanam, ``A time series is worth 64 words: Long-term forecasting with transformers,'' \emph{arXiv preprint arXiv:2211.14730}, 2022.

\bibitem{ngai2022emotion}
W.~K. Ngai, H.~Xie, D.~Zou, and K.-L. Chou, ``Emotion recognition based on convolutional neural networks and heterogeneous bio-signal data sources,'' \emph{Information Fusion}, vol.~77, pp. 107--117, 2022.

\bibitem{tao2020eeg}
W.~Tao, C.~Li, R.~Song, J.~Cheng, Y.~Liu, F.~Wan, and X.~Chen, ``Eeg-based emotion recognition via channel-wise attention and self attention,'' \emph{IEEE Transactions on Affective Computing}, vol.~14, no.~1, pp. 382--393, 2020.

\bibitem{li2022eeg}
C.~Li, X.~Lin, Y.~Liu, R.~Song, J.~Cheng, and X.~Chen, ``Eeg-based emotion recognition via efficient convolutional neural network and contrastive learning,'' \emph{IEEE Sensors Journal}, vol.~22, no.~20, pp. 19\,608--19\,619, 2022.

\bibitem{liu20213dcann}
S.~Liu, X.~Wang, L.~Zhao, B.~Li, W.~Hu, J.~Yu, and Y.-D. Zhang, ``3dcann: A spatio-temporal convolution attention neural network for eeg emotion recognition,'' \emph{IEEE Journal of Biomedical and Health Informatics}, vol.~26, no.~11, pp. 5321--5331, 2021.

\bibitem{lin2023eeg}
X.~Lin, J.~Chen, W.~Ma, W.~Tang, and Y.~Wang, ``Eeg emotion recognition using improved graph neural network with channel selection,'' \emph{Computer Methods and Programs in Biomedicine}, vol. 231, p. 107380, 2023.

\bibitem{cui2020eeg}
H.~Cui, A.~Liu, X.~Zhang, X.~Chen, K.~Wang, and X.~Chen, ``Eeg-based emotion recognition using an end-to-end regional-asymmetric convolutional neural network,'' \emph{Knowledge-Based Systems}, vol. 205, p. 106243, 2020.

\bibitem{deng2021sfe}
X.~Deng, J.~Zhu, and S.~Yang, ``Sfe-net: Eeg-based emotion recognition with symmetrical spatial feature extraction,'' in \emph{Proceedings of the 29th ACM international conference on multimedia}, 2021, pp. 2391--2400.

\bibitem{gong2023astdf}
P.~Gong, Z.~Jia, P.~Wang, Y.~Zhou, and D.~Zhang, ``Astdf-net: Attention-based spatial-temporal dual-stream fusion network for eeg-based emotion recognition,'' in \emph{Proceedings of the 31st ACM International Conference on Multimedia}, 2023, pp. 883--892.

\bibitem{cheng2023hybrid}
C.~Cheng, Z.~Yu, Y.~Zhang, and L.~Feng, ``Hybrid network using dynamic graph convolution and temporal self-attention for eeg-based emotion recognition,'' \emph{IEEE Transactions on Neural Networks and Learning Systems}, 2023.

\bibitem{gong2023eeg}
L.~Gong, M.~Li, T.~Zhang, and W.~Chen, ``Eeg emotion recognition using attention-based convolutional transformer neural network,'' \emph{Biomedical Signal Processing and Control}, vol.~84, p. 104835, 2023.

\bibitem{shen2022multiscale}
L.~Shen, M.~Sun, Q.~Li, B.~Li, Z.~Pan, and J.~Lei, ``Multiscale temporal self-attention and dynamical graph convolution hybrid network for eeg-based stereogram recognition,'' \emph{IEEE Transactions on Neural Systems and Rehabilitation Engineering}, vol.~30, pp. 1191--1202, 2022.

\bibitem{dosovitskiy2020image}
A.~Dosovitskiy, L.~Beyer, A.~Kolesnikov, D.~Weissenborn, X.~Zhai, T.~Unterthiner, M.~Dehghani, M.~Minderer, G.~Heigold, S.~Gelly \emph{et~al.}, ``An image is worth 16x16 words: Transformers for image recognition at scale,'' \emph{arXiv preprint arXiv:2010.11929}, 2020.

\bibitem{zhu2024vision}
L.~Zhu, B.~Liao, Q.~Zhang, X.~Wang, W.~Liu, and X.~Wang, ``Vision mamba: Efficient visual representation learning with bidirectional state space model,'' \emph{arXiv preprint arXiv:2401.09417}, 2024.

\bibitem{devlin2018bert}
J.~Devlin, M.-W. Chang, K.~Lee, and K.~Toutanova, ``Bert: Pre-training of deep bidirectional transformers for language understanding,'' \emph{arXiv preprint arXiv:1810.04805}, 2018.

\bibitem{schuster2012japanese}
M.~Schuster and K.~Nakajima, ``Japanese and korean voice search,'' in \emph{2012 IEEE international conference on acoustics, speech and signal processing (ICASSP)}.\hskip 1em plus 0.5em minus 0.4em\relax IEEE, 2012, pp. 5149--5152.

\bibitem{wang2024timemixer}
S.~Wang, H.~Wu, X.~Shi, T.~Hu, H.~Luo, L.~Ma, J.~Y. Zhang, and J.~Zhou, ``Timemixer: Decomposable multiscale mixing for time series forecasting,'' \emph{arXiv preprint arXiv:2405.14616}, 2024.

\bibitem{wu2022timesnet}
H.~Wu, T.~Hu, Y.~Liu, H.~Zhou, J.~Wang, and M.~Long, ``Timesnet: Temporal 2d-variation modeling for general time series analysis,'' \emph{arXiv preprint arXiv:2210.02186}, 2022.

\bibitem{chen2024pathformer}
P.~Chen, Y.~Zhang, Y.~Cheng, Y.~Shu, Y.~Wang, Q.~Wen, B.~Yang, and C.~Guo, ``Pathformer: Multi-scale transformers with adaptive pathways for time series forecasting,'' \emph{arXiv preprint arXiv:2402.05956}, 2024.

\bibitem{wang2020linking}
H.~Wang, L.~Xu, A.~Bezerianos, C.~Chen, and Z.~Zhang, ``Linking attention-based multiscale cnn with dynamical gcn for driving fatigue detection,'' \emph{IEEE Transactions on Instrumentation and Measurement}, vol.~70, pp. 1--11, 2020.

\bibitem{jiang2023emotion}
Y.~Jiang, S.~Xie, X.~Xie, Y.~Cui, and H.~Tang, ``Emotion recognition via multiscale feature fusion network and attention mechanism,'' \emph{IEEE Sensors Journal}, vol.~23, no.~10, pp. 10\,790--10\,800, 2023.

\bibitem{li2020multi}
D.~Li, J.~Xu, J.~Wang, X.~Fang, and Y.~Ji, ``A multi-scale fusion convolutional neural network based on attention mechanism for the visualization analysis of eeg signals decoding,'' \emph{IEEE Transactions on Neural Systems and Rehabilitation Engineering}, vol.~28, no.~12, pp. 2615--2626, 2020.

\bibitem{bai2020adaptive}
L.~Bai, L.~Yao, C.~Li, X.~Wang, and C.~Wang, ``Adaptive graph convolutional recurrent network for traffic forecasting,'' \emph{Advances in neural information processing systems}, vol.~33, pp. 17\,804--17\,815, 2020.

\bibitem{sesti2021integrating}
N.~Sesti, J.~J. Garau-Luis, E.~Crawley, and B.~Cameron, ``Integrating lstms and gnns for covid-19 forecasting,'' \emph{arXiv preprint arXiv:2108.10052}, 2021.

\bibitem{grigsby2021long}
J.~Grigsby, Z.~Wang, N.~Nguyen, and Y.~Qi, ``Long-range transformers for dynamic spatiotemporal forecasting,'' \emph{arXiv preprint arXiv:2109.12218}, 2021.

\bibitem{chen2020low}
X.~Chen and L.~Sun, ``Low-rank autoregressive tensor completion for multivariate time series forecasting,'' \emph{arXiv preprint arXiv:2006.10436}, 2020.

\bibitem{sharma2022classification}
A.~Sharma and D.~Kumar, ``Classification with 2-d convolutional neural networks for breast cancer diagnosis,'' \emph{Scientific Reports}, vol.~12, no.~1, p. 21857, 2022.

\bibitem{jin2023graph}
M.~Jin and J.~Li, ``Graph to grid: Learning deep representations for multimodal emotion recognition,'' in \emph{Proceedings of the 31st ACM International Conference on Multimedia}, 2023, pp. 5985--5993.

\bibitem{li2024multi}
C.~Li, N.~Bian, Z.~Zhao, H.~Wang, and B.~W. Schuller, ``Multi-view domain-adaptive representation learning for eeg-based emotion recognition,'' \emph{Information Fusion}, vol. 104, p. 102156, 2024.

\bibitem{cheng2024dense}
C.~Cheng, W.~Liu, L.~Feng, and Z.~Jia, ``Dense graph convolutional with joint cross-attention network for multimodal emotion recognition,'' \emph{IEEE Transactions on Computational Social Systems}, 2024.

\bibitem{zeng2023transformers}
A.~Zeng, M.~Chen, L.~Zhang, and Q.~Xu, ``Are transformers effective for time series forecasting?'' in \emph{Proceedings of the AAAI conference on artificial intelligence}, vol.~37, no.~9, 2023, pp. 11\,121--11\,128.

\bibitem{liu2023itransformer}
Y.~Liu, T.~Hu, H.~Zhang, H.~Wu, S.~Wang, L.~Ma, and M.~Long, ``itransformer: Inverted transformers are effective for time series forecasting,'' \emph{arXiv preprint arXiv:2310.06625}, 2023.

\bibitem{koelstra2011deap}
S.~Koelstra, C.~Muhl, M.~Soleymani, J.-S. Lee, A.~Yazdani, T.~Ebrahimi, T.~Pun, A.~Nijholt, and I.~Patras, ``Deap: A database for emotion analysis; using physiological signals,'' \emph{IEEE transactions on affective computing}, vol.~3, no.~1, pp. 18--31, 2011.

\bibitem{katsigiannis2017dreamer}
S.~Katsigiannis and N.~Ramzan, ``Dreamer: A database for emotion recognition through eeg and ecg signals from wireless low-cost off-the-shelf devices,'' \emph{IEEE journal of biomedical and health informatics}, vol.~22, no.~1, pp. 98--107, 2017.

\bibitem{zheng2015investigating}
W.-L. Zheng and B.-L. Lu, ``Investigating critical frequency bands and channels for eeg-based emotion recognition with deep neural networks,'' \emph{IEEE Transactions on autonomous mental development}, vol.~7, no.~3, pp. 162--175, 2015.

\bibitem{liu2022non}
Y.~Liu, H.~Wu, J.~Wang, and M.~Long, ``Non-stationary transformers: Exploring the stationarity in time series forecasting,'' \emph{Advances in Neural Information Processing Systems}, vol.~35, pp. 9881--9893, 2022.

\bibitem{zhou2021informer}
H.~Zhou, S.~Zhang, J.~Peng, S.~Zhang, J.~Li, H.~Xiong, and W.~Zhang, ``Informer: Beyond efficient transformer for long sequence time-series forecasting,'' in \emph{Proceedings of the AAAI conference on artificial intelligence}, vol.~35, no.~12, 2021, pp. 11\,106--11\,115.

\bibitem{bai2018empirical}
S.~Bai, J.~Z. Kolter, and V.~Koltun, ``An empirical evaluation of generic convolutional and recurrent networks for sequence modeling,'' \emph{arXiv preprint arXiv:1803.01271}, 2018.

\bibitem{li2022multi}
C.~Li, Y.~Hou, R.~Song, J.~Cheng, Y.~Liu, and X.~Chen, ``Multi-channel eeg-based emotion recognition in the presence of noisy labels,'' \emph{Science China Information Sciences}, vol.~65, no.~4, p. 140405, 2022.

\bibitem{song2020instance}
T.~Song, S.~Liu, W.~Zheng, Y.~Zong, and Z.~Cui, ``Instance-adaptive graph for eeg emotion recognition,'' in \emph{Proceedings of the AAAI Conference on Artificial Intelligence}, vol.~34, no.~03, 2020, pp. 2701--2708.

\bibitem{song2021variational}
T.~Song, S.~Liu, W.~Zheng, Y.~Zong, Z.~Cui, Y.~Li, and X.~Zhou, ``Variational instance-adaptive graph for eeg emotion recognition,'' \emph{IEEE Transactions on Affective Computing}, vol.~14, no.~1, pp. 343--356, 2021.

\bibitem{xu2024eescn}
F.~Xu, D.~Pan, H.~Zheng, Y.~Ouyang, Z.~Jia, and H.~Zeng, ``Eescn: A novel spiking neural network method for eeg-based emotion recognition,'' \emph{Computer methods and programs in biomedicine}, vol. 243, p. 107927, 2024.

\bibitem{cheng2024novel}
C.~Cheng, W.~Liu, Z.~Fan, L.~Feng, and Z.~Jia, ``A novel transformer autoencoder for multi-modal emotion recognition with incomplete data,'' \emph{Neural Networks}, vol. 172, p. 106111, 2024.

\end{thebibliography}
\end{document}